\documentclass[12pt]{article}  
\usepackage{graphicx}
\def \bea{\begin{eqnarray}}
\def \beq{\begin{equation}}
\def \bo{B^0}
\def \eea{\end{eqnarray}}
\def \eeq{\end{equation}}

\def \e1{\hb \to \gamma \eb}
\def \eb{\eta_b}
\def \hb{h_b}
\def \ko{K^0}
\def \ob{\overline{B}^0}
\def \ok{\overline{K}^0}

\begin{document}
\renewcommand{\thetable}{\Roman{table}}
\begin{flushright}
EFI 06-21 \\
hep-ph/0610100 \\
October, 2006 \\
\end{flushright}

\centerline{\bf UNCOVERING THE NATURE OF THE WEAK INTERACTION
\footnote{Invited talk presented at Jim Cronin's 75 birthday celebration,
Chicago, September 8--9, 2006}}
\medskip

\centerline{Jonathan L. Rosner}
\centerline{\it Enrico Fermi Institute and Department of Physics, University
of Chicago}
\centerline{\it 5640 South Ellis Avenue, Chicago, IL 60637}

\begin{quote}
This brief review traces the development of our understanding of the weak
interactions, highlighting Jim Cronin's contributions through his studies
of strange particle decays and the developments to which they led.
\end{quote}

\section{INTRODUCTION}

The history of the weak interactions may be said to have begun with the
discovery of beta-decay by Henri Becquerel at the end of the 19th Century
\cite{Becq}.  It is still evolving.  Insights are expected from experiments
ranging from neutrinoless double beta-decay to giant cosmic ray air showers.
It has benefitted greatly from Jim Cronin's studies of strange particles,
including but not limited to his discovery of CP violation with Christenson,
Fitch, and Turlay \cite{Christenson:1964fg}.

In a thirty-minute talk (or a written version thereof) it is impossible to do
justice to this rich 110-year history.  Jim entered the world roughly nine
months after Pauli's December 4, 1930 proposal of the neutrino \cite{Pauli},
but was already a practicing physicist when its discovery was announced
\cite{neutrino}.  I would like to touch upon some high points, paying special
heed to the term ``uncovering.'' Jim has been on the front-line of this
effort.  The fundamental weak interactions often have been overlaid with 
strong interactions and kinematic correlations, whose understanding is needed
to draw conclusions about the underlying physics.  For example:

\begin{enumerate}

\item In nuclear beta-decay, $0 \to 0$ (``Fermi'') transitions have proven
especially simple to describe.  Complications of nuclear matrix elements,
often plaguing interpretations, are at a minimum for these transitions.

\item In the discovery that the weak interactions violated parity conservation,
a key role was played by Dalitz's ``phase space plots'' \cite{Dalitz:1953cp}.

\item The nucleon axial-vector coupling, a pure number differing from unity,
was related to strong-interaction parameters by Goldberger and Treiman
\cite{Goldberger:1958tr}.

\item Nonleptonic hyperon decays to a pion and a baryon can proceed in general
both through parity-violating S-wave and parity-conserving P-wave decays.  The
interference of these two amplitudes can lead to decay asymmetries and provide
tests of time-reversal invariance.

\item The comparison of $K_{\mu 3}$ and $K_{e3}$ semileptonic decays provides
a test of lepton universality in the weak interactions, but also allows one
to distinguish two independent form factors in the $K \to \pi$ weak
transition.  These decays provide fundamental information about
strangeness-changing weak decays once such questions are resolved.

\item Hyperon semileptonic beta-decays \cite{Cabibbo:1963yz} also give
useful information about strangeness-changing weak decays once one
understands patterns of baryonic matrix elements of the weak current.

\item The interpretation of the phase of the CP-violating amplitude in neutral
kaon decays requires one to understand S-wave pion-pion scattering at a
center-of-mass energy equal to the neutral kaon mass.

\item The algebra of currents \cite{Gell-Mann:1962xb} extracted some essential
features of quarks without the need for them to be real entities.  It
pointed the way to the necessary features of a strong-interaction theory.

\item The charmed quark played a key role in unifying the weak and
electromagnetic interactions.  Interpretation of initial evidence for it
was greatly facilitated by the emerging theory of the strong interactions,
quantum chromodynamics (QCD).

\item A candidate theory of CP violation, proposed by Kobayashi and Maskawa
more than thirty years ago \cite{Kobayashi:1973fv}, has passed all tests so
far with flying colors.  Some of these tests, using mesons containing the
fifth (``$b$'') quark, require one to separate strong-interaction and
weak-interaction effects; others are unaffected by strong interactions.
As in so many of the above cases, the trick lies in recognizing which
measurements yield the most fundamental quantities.

\end{enumerate}

There are many more such examples.  We shall discuss items \#4 (nonleptonic
hyperon decays), \#6 (semileptonic hyperon decays), \#9 (charm), and \#10
(the Kobayashi-Maskawa theory) in subsequent sections.

\section{NONLEPTONIC HYPERON DECAYS}
\medskip

When the weak interactions were shown to violate parity conservation, a
natural expectation was the expectation that hyperon decays would also
violate parity \cite{Gell-Mann:1957wh}.  By the late 1950s, parity violation
had been seen in polarized $\Lambda$ decays, manifested by an up-down asymmetry
of protons in $\Lambda \to \pi^- p$ produced in $\pi^-p \to \Lambda K^0$.  Did
it occur in $\Sigma$ decays?  Jim Cronin played a key role in sorting out
such decays as $\Sigma^+ \to \pi^+ n$, $\Sigma^+ \to \pi^0 p$, and $\Sigma^-
\to \pi^- n$ \cite{Cool:1959} and performing tests of time-reversal violation
in $\Lambda \to \pi^- p$ \cite{Cronin:1963zb}.

The amplitudes for transitions $(J^P = 1/2^+) \to (J^P=0^-) + (J^P=1/2^+)$,
where $J$ denotes total angular momentum and $P$ denotes parity, can be S-wave 
(``$s$,'' parity-violating) or P-wave (``$p$,'' parity-conserving).  Decays
are fully characterized by the parameters
\bea
\Gamma & \sim & |s|^2 + |p|^2~,~
        \alpha \equiv 2{\rm~Re}(s p^*)/(|s|^2 + |p|^2)~,~
         \beta \equiv 2{\rm~Im}(s p^*)/(|s|^2 + |p|^2)~,\\
\gamma & \equiv & (|s|^2 - |p|^2)/(|s|^2 + |p|^2)~,~~~
        \alpha^2 + \beta^2 + \gamma^2 = 1~.
\eea
The parameter $\beta$ is a coefficient of a T-violating observable in the
decay, and is expected to have a specific non-zero value as a result of
final-state interactions.

\subsection{$\Sigma \to \pi N$ and its interpretation}

In Ref.\ \cite{Cool:1959} Cronin and his collaborators measured $\alpha
{\cal P}$, where ${\cal P}$ denotes the hyperon polarization, via the
$\Sigma^\pm$ decay asymmetry with respect to the plane formed by the incident
beam and the recoiling hyperon, for example in $\pi^- p \to \Sigma^- K^+$
or $\pi^+ p \to \Sigma^+ K^+$.  They found $\alpha(\Sigma^+ \to \pi^0 p)
{\cal P}(\Sigma^+) = 0.70 \pm 0.30$, with $\alpha(\Sigma^+ \to \pi^+n) {\cal P}
(\Sigma^+) = 0.02 \pm 0.07$ for an initial beam momentum of 1 GeV/$c$, and
$\alpha(\Sigma^- \to n \pi^-) {\cal P}(\Sigma^-)$ consistent with 0 at beam
momenta 1 and 1.1 GeV/$c$.  They thus concluded that parity violation is large
in $\Sigma^+ \to \pi^0 p$.  However, they were able to extend the implications
of their measurements considerably with the following observations:
(1) The rates for all three $\Sigma \to \pi N$ decays are nearly equal; (2) The
nonleptonic weak interaction greatly favors transitions with $\Delta I = 1/2$
over those with $\Delta I = 3/2$.

\begin{figure}
\begin{center}
\includegraphics[height=3.0in]{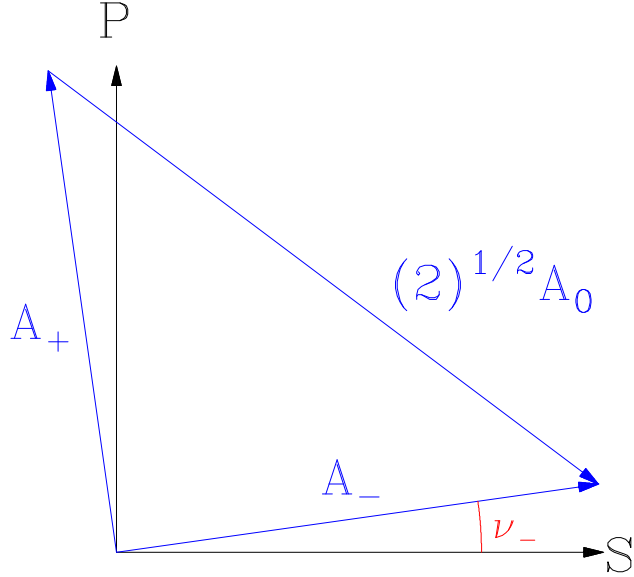}
\end{center}
\caption{Triangle formed by amplitudes for $\Sigma \to \pi N$ decays.
\label{fig:tri}}
\end{figure}

The $\Delta I = 1/2$ rule implies $A_+ + \sqrt{2} A_0 = A_-$, where the
subscript denotes the pion charge, so the amplitudes $A_\pm$ and $\sqrt{2} A_0$
form an isosceles right triangle shown in Fig.\ \ref{fig:tri}, where the axes
denote S-wave and P-wave amplitudes.

% This is Figure 1

One can show that the decay asymmetry parameters $\alpha$ for the three decays
are related to one another by $\alpha^- = - \alpha^+ \equiv \sin 2 \nu_-$,
$\alpha^0 = \pm \cos 2 \nu_-$, where the superscript denotes the pion charge.
The $\pm$ sign occurs because the triangle
could have been drawn reflected about $A_-$.  In the context of equal rates for
all $\Sigma \to \pi N$ decays and the $\Delta I = 1/2$ rule, Cronin and his
colleagues interpreted the data to imply
\beq
\alpha^+ = - \alpha^- \le \pm(0.03 \pm 0.11)~,~~
\alpha^0 = \pm (0.99 \pm 0.01)~.
\eeq
The current (2006) values \cite{Yao:2006px} are  $\alpha^+ = 0.068 \pm 0.013$,
$\alpha^- = -0.068 \pm 0.008$, $\alpha^0 = -0.980^{+0.017}_{-0.015}$, very
close to those measured nearly forty years ago.

\subsection{Measurement of $(\alpha,\beta)$ in $\Lambda \to \pi^- p$}

Measurements of an up-down asymmetry with respect to a production plane, such
as those just described, provide only the product $\alpha {\cal P}$ of the
asymmetry parameter and the polarization.  To measure $\alpha$ separately one
needs the polarization of the final baryon, for example that of the proton in
$\Lambda \to \pi^- p$.

Cronin and Overseth \cite{Cronin:1963zb} measured ${\cal P}(p)$ from scattering
of the final proton in carbon plates.  (See Ref.\ \cite{Engels:2006} for
instrumental details.)  They found $\alpha = 0.62 \pm 0.07$, to be compared
with the present value of $0.642 \pm 0.013$ \cite{Yao:2006px}, and $\beta =
0.18 \pm 0.24$, to be compared with the present value tan$^{-1}(\beta/\alpha) =
(8 \pm 4)^\circ$) \cite{Yao:2006px}.  From these measurements and one of
$\gamma = 0.78 \pm 0.06$ they concluded that the $|p/s|$ ratio was small.  When
combined with
information on hypernuclei, this allowed them to conclude that the relative $K
\Lambda N$ parity was odd.  Thus, if the $K$ and $\pi$ both had the same (odd)
parity, as predicted if they belonged to the same SU(3) multiplet, the proton
and $\Lambda$ would also have the same (even) parity, in accord with their
assignment to the same SU(3) octet.

\section{STRANGE PARTICLE SEMILEPTONIC DECAYS}

Strangeness-changing $|\Delta S| = 1$ weak semileptonic decays were seen to be
suppressed in comparison with those having $\Delta S = 0$.  This led
Gell-Mann and L\'evy \cite{Gell-Mann:1960np} to propose a weak current
taking $p \leftrightarrow n \cos \theta + \Lambda \sin \theta$.  Cabibbo
\cite{Cabibbo:1963yz} generalized this; in quark language his form of the
transition reads $u \leftrightarrow d \cos \theta + s \sin \theta$.  His
approach used SU(3) symmetry to relate matrix elements of the weak current to
one another.

\begin{figure}
\mbox{\includegraphics[height=1.75in]{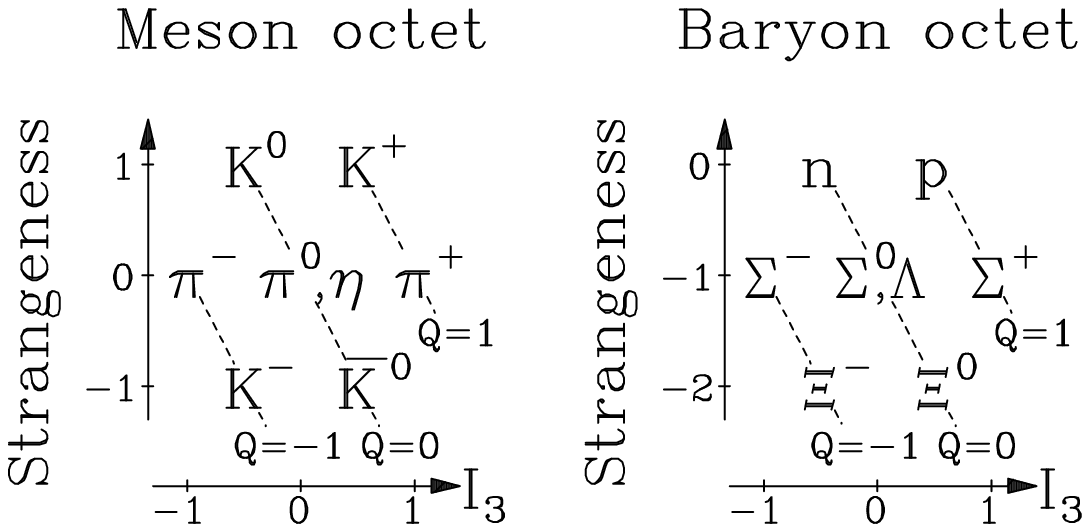} \hskip 0.2in
      \includegraphics[height=1.8in]{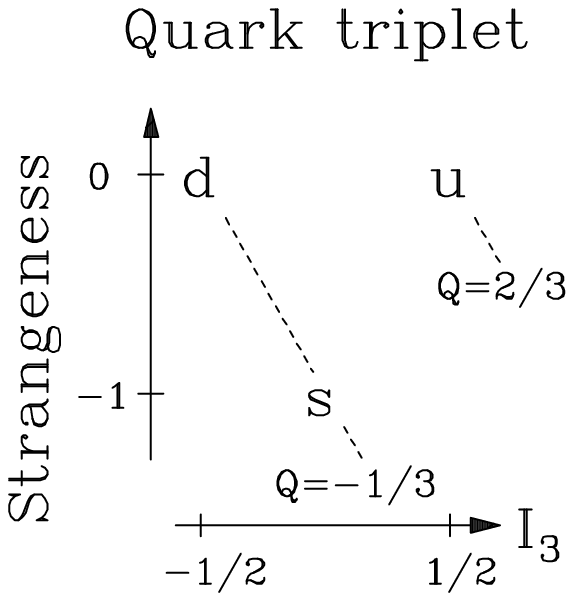}}
\caption{Plots of strangeness vs.\ third component of isotopic spin ($I_3$)
for meson and baryon octets and quarks.
\label{fig:i3y}}
\end{figure}

The mesons and baryons to which Cabibbo's proposal applied are
shown in Fig.\ \ref{fig:i3y}, along with the lightest three quarks.  The
proposal implied $\Delta S = \Delta Q$ in weak semileptonic decays and
successfully described such transitions as $n \to p e^- \bar \nu_e$, $\Lambda
\to p e^- \bar \nu_e$, $\Sigma^- \to \Lambda e^- \bar \nu_e$, $\Sigma^- \to n
e^- \bar \nu_e$, $\Xi^- \to (\Sigma^0,\Lambda) e^- \bar \nu_e$, $\Xi^0 \to
\Sigma^+ e^- \bar \nu_e$, $K^- \to \pi^0 e^- \bar \nu_e$, and
$\bar K^0 \to \pi^+
e^- \nu_e$.  The latest average \cite{Yao:2006px} (including a key Chicago
measurement \cite{Alexopoulos:2004sw}) is $\sin \theta = 0.2257 \pm 0.0021$.
The Chicago contribution relied on an understanding of form factors in
$K_{e3}$ and $K_{\mu 3}$ decay \cite{Alexopoulos:2004sy} which has a long
history including important early contributions by Jim Cronin and collaborators
at Saclay \cite{Basile:1970hj}.  Cronin's reviews of the weak interactions
in the late 1960s give a good picture of the emerging success of the
Cabibbo theory \cite{Cronin:1967}.

% This is Figure 2

\section{WEAK LEPTONIC, HADRONIC CURRENTS}

The Fermi interaction (including parity violation) was described
by a Hamiltonian density
\beq
{\cal H}_W = (G_F/\sqrt{2})(J^{\rm lepton} + J^{\rm hadron})_\alpha
(J^{\dag~{\rm lepton}} + J^{\dag~{\rm hadron}})^\alpha~~.
\eeq
Since 1962 it was known that each lepton $e^-, \mu^-$ had its own neutrino
$\nu_e,\nu_\mu$, so the weak charge-changing current of leptons could be
written
\beq
J^{\rm lepton}_\alpha = \bar \nu_e \gamma_\alpha (1 - \gamma_5) e 
 + \bar \nu_\mu \gamma_\alpha (1 - \gamma_5) \mu~~.
\eeq
The spatial integral $Q^{(+)} = \frac{1}{2} \int d^3x J^{\rm lepton}_0$ of its
time-component satisfies {\it commutation relations} with its Hermitian adjoint
$Q^{(-)} = Q^{(+)\dag}$:
\beq \label{eqn:cr}
Q_3 \equiv \frac{1}{2}[Q^{(+)},Q^{(-)}],~[Q_3,Q^{(\pm)}] = \pm Q^{(\pm)}~~.
\eeq
These are just the commutation relations of SU(2), with
$Q^{(+)}$ acting as a ``raising operator,'' and serve to normalize the
leptonic current.  Gell-Mann (1962) proposed similar commutation relations
[for SU(3) and vector, axial currents] to normalize the weak currents of {\it
hadrons}.  With $J^{\rm hadron}_\alpha = \bar u \gamma_\alpha (1 -
\gamma_5) (d \cos \theta + s \sin \theta)$, $Q_3^{\rm hadron}$ has terms
inducing $s \leftrightarrow d$.  These are harmless if $Q_3$ doesn't couple to
anything.  However, the electroweak theory of Glashow \cite{Glashow:1961tr},
Weinberg \cite{Weinberg:1967tq}, and Salam \cite{Salam:1968} implied that
it {\it does}, in contradiction to experiment.  For this reason Weinberg
entitled his model ``A Theory of Leptons.''
 
\subsection{From Fermi to the electroweak theory}

% This is Figure 3
\begin{figure}
\mbox{\includegraphics[height=2.1in]{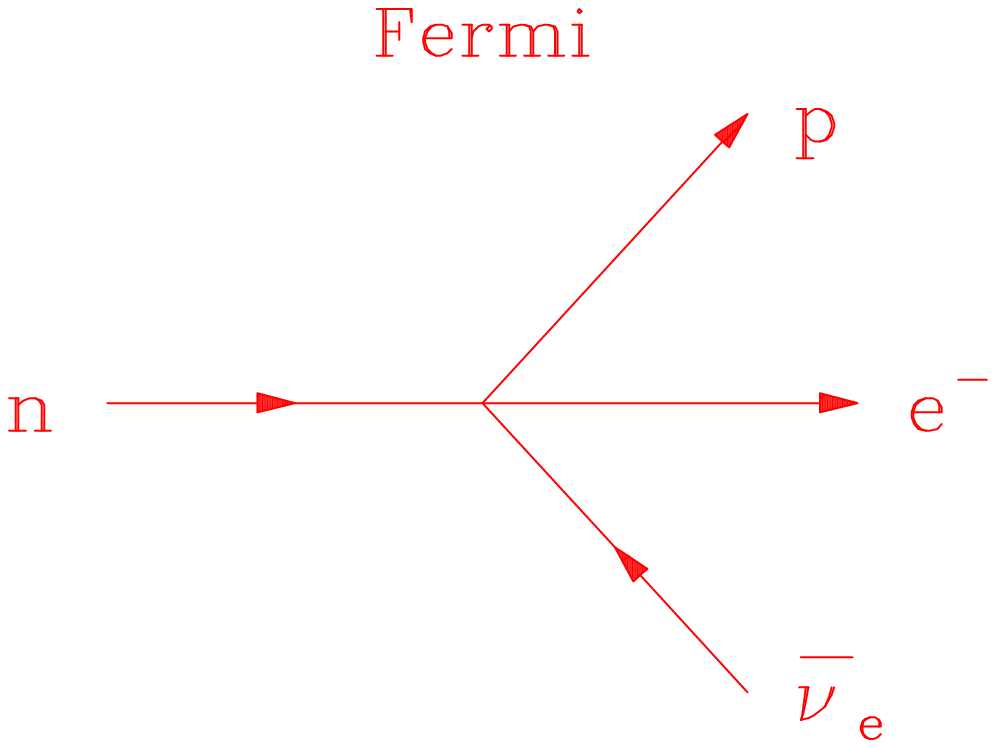} \hskip 0.2in
 \includegraphics[height=2.2in]{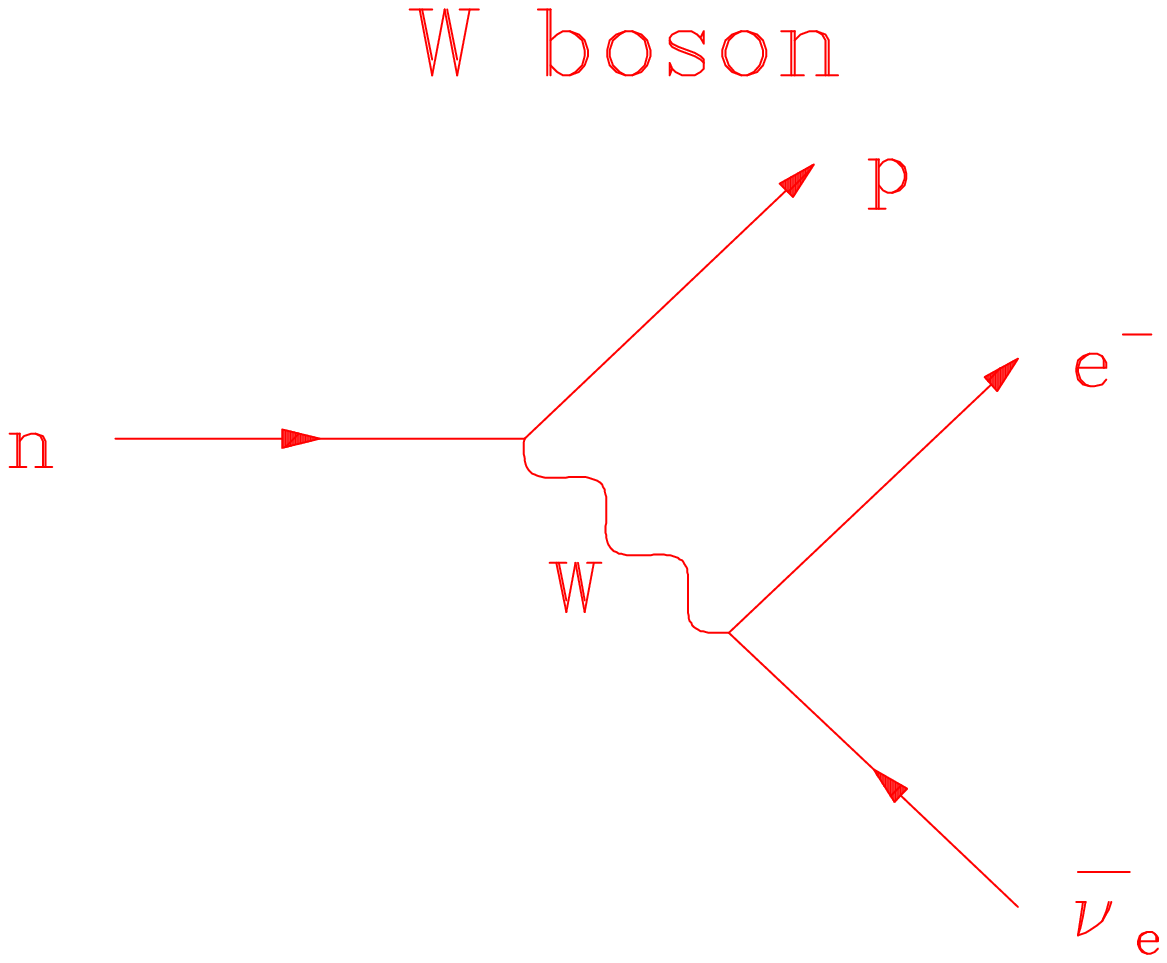}}
\caption{Fermi (left) and $W$ boson (right) pictures of
charge-changing weak interactions.
\label{fig:fgws}}
\end{figure}

Many authors, starting with Yukawa, in order to avoid the singular 4-fermion
interaction of the Fermi theory, proposed that weak interactions were due to
exchange of an intermediate boson $W$ \cite{Wboson}, as shown in Fig.\
\ref{fig:fgws}.  Charged $W$'s would be members $(W_1 \pm i W_2)/\sqrt{2}$
of an SU(2) triplet.  However, the photon could not be the neutral member; its
coupling doesn't violate parity.  To solve this problem, Glashow
\cite{Glashow:1961tr} proposed an extension to SU(2) $\times$ U(1), implying
the existence of an additional neutral boson $Z$ with $M_Z > M_W$.  The photon
and $Z$ are then orthogonal mixtures of $W_3$ and the U(1) boson $B$.  To break
the SU(2) $\times$ U(1) symmetry, Weinberg \cite{Weinberg:1967tq} and Salam
\cite{Salam:1968} utilized the Higgs mechanism, entailing the existence of a
spinless boson which is still the subject of intense searches.

\subsection{Neutral currents and charm}

The leptonic $Q_3$ calculated from Eq.\ (\ref{eqn:cr}) takes $e \leftrightarrow
e$, $\mu \leftrightarrow \mu$, $\nu_e \leftrightarrow \nu_e$, $\nu_\mu
\leftrightarrow \nu_\mu$, i.e., the neutral current is {\it flavor-preserving}.
Can this be arranged for quarks?  By pursuing a quark-lepton analogy, Bjorken
and Glashow \cite{Bjorken:1964qz}, Hara \cite{Hara:1963gw}, and Maki and Ohnuki
\cite{Maki:1964} introduced a second quark $c$ (``charm'') with charge $Q=2/3$
so that the charge-changing weak hadron current became
\beq
J^{\rm hadron}_\alpha
= \bar u \gamma_\alpha (1 - \gamma_5) (d \cos \theta + s \sin \theta)
+ \bar c \gamma_\alpha (1 - \gamma_5) (-d \sin \theta + s \cos \theta)
\eeq
Calculating $Q_3$ for hadrons, one finds that it takes $u \leftrightarrow u$,
$c \leftrightarrow c$, $d \leftrightarrow d$, and $s \leftrightarrow s$, i.e.,
it has {\it no flavor-changing neutral currents}.  There are thus two families
of quarks and leptons, with an orthogonal mixing matrix for quarks as shown in
Table \ref{tab:4qk}.

% This is Table I
\begin{table}
\caption{Pattern of couplings in the four-quark theory involving the quarks
$u,~d,~c,~s$
\label{tab:4qk}}
\begin{center}
\begin{tabular}{|c|c|c|} \hline
Leptons & Quarks & Quark mixing \\ \hline
$\left( \begin{array}{c} \nu_e \\ e^- \end{array} \right)$
$\left( \begin{array}{c} \nu_\mu \\ \mu^- \end{array} \right)$ &
$\left( \begin{array}{c} u \\ d' \end{array} \right)$
$\left( \begin{array}{c} c \\ s' \end{array} \right)$ &
$\left( \begin{array}{c} d' \\ s' \end{array} \right)$ =
$\left( \begin{array}{c c} \cos \theta & \sin \theta \\
 - \sin \theta & \cos \theta \end{array} \right)$
$\left( \begin{array}{c} d \\ s \end{array} \right)$ \\ \hline
\end{tabular}
\end{center}
\end{table}

In 1970, Glashow, Iliopoulos, and Maiani \cite{Glashow:1970gm} showed that the
charmed quark suppressed flavor-changing neutral currents which are induced in
higher-order calculations (such as $\ko$--$\ok$ mixing).  In the context of the
newly developed electroweak theory, Gaillard and Lee \cite{Gaillard:1974hs}
demonstrated cancellations due to charm in many rare kaon decays, and estimated
that the charmed quark could have a mass of no more than 2 GeV/$c^2$.  At a
conference in the spring of 1974, Glashow offered to eat his hat if charm hadn't
been discovered by the next conference in the series \cite{Glashow:1974ru}.

Hints of charm were already emerging at the London (1974) International
Conference on High Energy Physics.  Jim Cronin and I discussed the anomalous
leptons observed at high transverse momenta, some of which turned out to be due
to charm decays, and Ben Lee remarked after a talk on neutrino interactions
that events with opposite-sign dimuons could be due to production and
subsequent semileptonic decay of charm.  Gaillard, Lee, and I analyzed various
experimental signatures of charm \cite{Gaillard:1974mw} in anticipation of its
imminent discovery.

\section{THE EMERGENCE OF CHARM}

The first hints of charm were provided by short tracks observed in nuclear
emulsion by K. Niu and his collaborators in 1971 \cite{Niu:1971xu}.  However,
the discovery which most physicists found convincing was the observation of a
narrow $^3S_1$ $c \bar c$ ground state called $J$ on the East Coast
\cite{Aubert:1974js} and $\psi$ on the West Coast \cite{Augustin:1974xw}.
The discovery of the first {\it charmonium} state (a bound state of a charmed
quark and its antiquark) not only validated the charm hypothesis but also
demonstrated the reality of quarks and the applicability of QCD to processes
involving heavy quarks and large momenta.

\begin{figure}
\begin{center}
\includegraphics[height=2in]{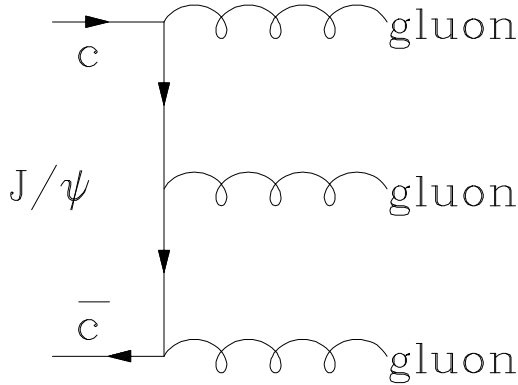}
\end{center}
\caption{Diagram for $J/\psi$ decay to three gluons.
\label{fig:3g}}
\end{figure}

QCD was developed as a strong-interaction theory which would preserve current
algebra:  It had to be a vector-like theory and to be {\it asymptotically
free} \cite{Gross:1973id,Politzer:1973fx}, with the weakening of interactions
at short distances allowing quarks to behave as quasi-free objects when probed
in deeply inelastic lepton scattering experiments.  Appelquist and Politzer
\cite{Appelquist:1974zd} used the newly developed QCD theory to predict that
the lowest $c \bar c$ $^3S_1$ state had a total width $\Gamma < 1$ MeV as a
result of the high order of pertubation theory needed to describe its decay
to light hadrons through three {\it gluons} (quanta of QCD), as shown in
Fig.\ \ref{fig:3g}.  The three-gluon width is proportional to $\alpha_s^3$,
where $\alpha_s$, the strong fine-structure constant, is about 0.3 at the scale
relevant for $J/\psi$ decay, and is further suppressed by the small 3-body
phase space.  A similar phase space suppression is responsible for the long
lifetime of orthopositronium.  The observed 3-gluon width is even smaller than
anticipated thanks to relativistic effects; the total width is $\Gamma_{\rm
tot}(J/\psi) \simeq 0.1$ MeV.

% This is Figure 4

\begin{figure}
\includegraphics[height=4in]{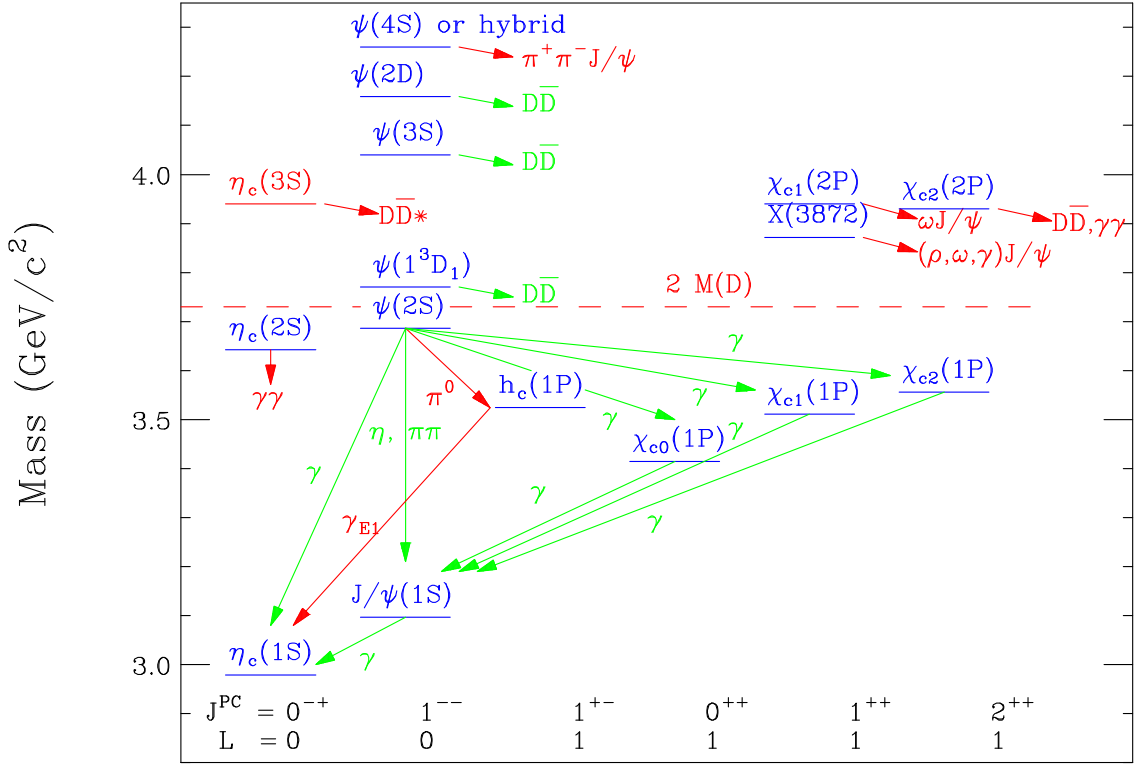}
\caption{Charmonium spectrum as of 2006.
\label{fig:charmon}}
\end{figure}

In the more than thiry years since its discovery, charmonium has evolved into a
fertile QCD laboratory.  There are now more known states for charmonium than
for positronium.  A snapshot of them is shown in Fig.\ \ref{fig:charmon}
\cite{Rosner:2006jz}.  The dark arrows denote transitions observed in the past
year or two by BaBar, Belle, CDF, and CLEO, and others.  The masses and
decays of these states serve as a test-bed for techniques of QCD, including
non-perturbative methods such as lattice gauge theory which are crucial for
tackling many long-distance properties of the states.

\section{KOBAYASHI-MASKAWA; CP VIOLATION}

Kobayashi and Maskawa \cite{Kobayashi:1973fv} took charm seriously.  They noted
that with only $(u,d)$ and $(c,s)$ one could always choose the charge-changing
couplings to be real, as in Table \ref{tab:4qk}.  With an additional pair of
quarks ($t$ for ``top'' and $b$ for ``bottom'' or ``beauty'') this was no
longer so; there emerged physically meaningful complex phases in couplings
describing charge-changing weak interactions, leading to CP violation.

% This is Figure 5

\begin{figure}
\begin{center}
\includegraphics[height=2.4in]{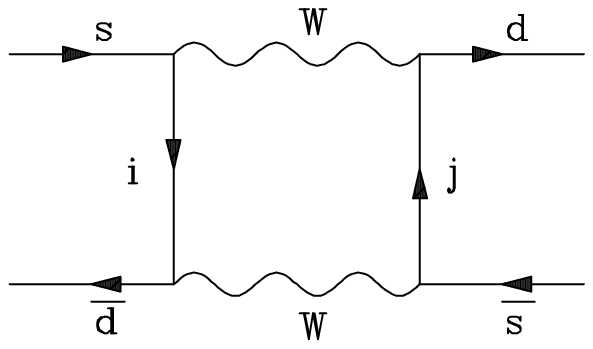}
\end{center}
\caption{Box diagram describing $\ko$--$\ok$ mixing.  Here $i,j = (u,c,t)$.
\label{fig:kmix}}
\end{figure}

The effect of the Kobayashi-Maskawa phases in neutral kaon decays mainly is to
induce CP-violating $K^0$--$\ok$ mixing through box diagrams dominated by the
heavy top quark contribution, as in Fig.\ \ref{fig:kmix}.  In one standard
parametrization \cite{Wolfenstein:1983yz} the key non-removable phase occurs
in the $t$-$d$-$W$ coupling $V_{td}$, so that CP-violating mixing arises with
a strength proportional to Im($V_{td}^2$).

% This is Figure 6

Key predictions of the Kobayashi-Maskawa theory of CP violation are:  (1) the
existence of the $b$ and $t$ quarks, verified by the discovery in 1977 of
the $\Upsilon$, a $b \bar b$ bound state, and its excitations
\cite{Herb:1977ek}, and in 1994 of the top quark $t$ \cite{Abe:1994xt}; (2)
direct CP violation in neutral kaon decay, affecting the ratio of CP-violating
to CP-conserving decays when comparing $K \to \pi^+ \pi^-$ with $K \to \pi^0
\pi^0$ \cite{Winstein:2006}; (3) large CP violation in $B$ meson decays
\cite{Carter:1980hr,Bigi:1981qs}.  The latter phenomenon has been the object
of recent studies with asymmetric $e^+ e^-$ colliders, an invention of
Pier Oddone to allow production of the $B$ mesons in a moving frame so that
their decays can be studied with greater resolution \cite{Oddone:1989ir}.
Experiments by the Belle Collaboration at KEK and the BaBar Collaboration at
SLAC have led to a wealth of information on $B$ decays, as we shall note
presently.

% This is Figure 7
\begin{figure}
% \begin{center}
\includegraphics[height=3.5in]{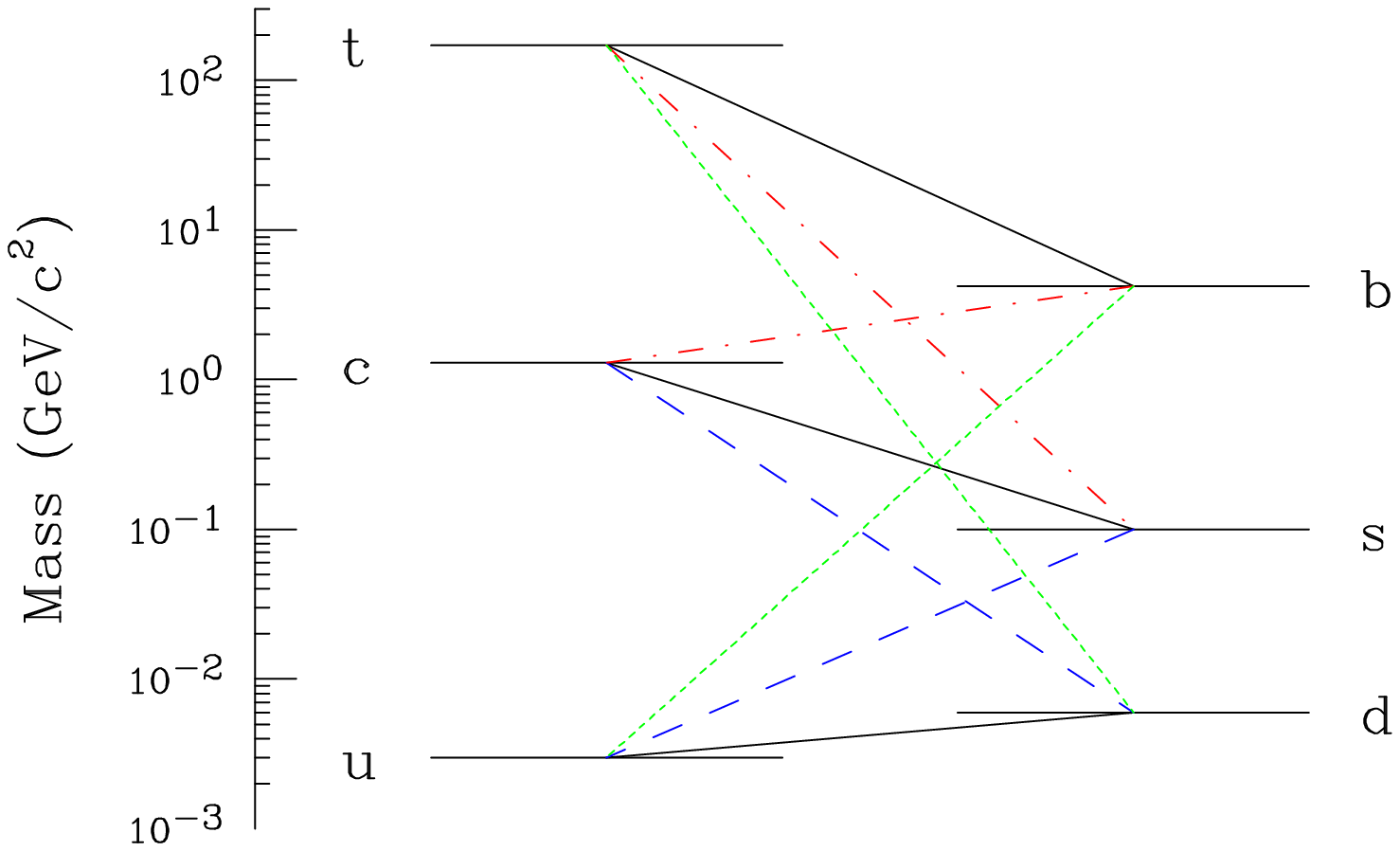}
% \end{center}
\caption{Quarks and the charge-changing weak transitions between them.
Relative strengths:  1 (solid), 0.22 (dashes), 0.04 (dotdash), $< 0.01$
(dotted).
\label{fig:qks}}
\end{figure}

The charge-changing weak transitions among the six quarks of the
Kobayashi-Maskawa theory are illustrated in Fig.\ \ref{fig:qks}.  The couplings
are described by a $3 \times 3$ unitary {\it Cabibbo-Kobayashi-Maskawa} (CKM)
matrix, a generalization of the $2 \times 2$ matrix illustrated in Table
\ref{tab:4qk}.  The approximate values of the CKM matrix elements are
$V_{ud} \simeq V_{cs} \simeq 0.974$, $V_{tb} \simeq 1$, $V_{us} \simeq -V_{cd}
\simeq 0.226$, $V_{cb} \simeq -V_{ts} \simeq 0.041$, $V_{td} \simeq 0.008
e^{-i~21^\circ},~V_{ub} \simeq 0.004e^{-i~60^\circ}$.  Here we have adopted a
parametrization invented by L. Wolfenstein \cite{Wolfenstein:1983yz}, in which
the large phases occur in $V_{td}$ and $V_{ub}$:

\beq
V = \left[ \begin{array}{c c c}
V_{ud} & V_{us} & V_{ub} \\
V_{cd} & V_{cs} & V_{cb} \\
V_{td} & V_{ts} & V_{tb} \end{array} \right] \simeq \left[ \begin{array}{c c c}
1 - \frac{\lambda^2}{2} & \lambda & A \lambda^3 (\rho - i \eta) \\
-\lambda & 1 - \frac{\lambda^2}{2} & A \lambda^2 \\
A \lambda^3 (1 - \rho - i \eta) & - A \lambda^2 & 1 \end{array} \right]
\eeq

No redefinition of the quark phases can get rid of all phases in $V$.  The
unitarity of this matrix ($V^\dag V = 1$) implies (e.g.) $V_{ud}V^*_{ub} +
V_{cd} V^*_{cb} + V_{td} V^*_{tb} = 0$ or (rescaling) $(\rho + i \eta) +
(1 - \rho - i \eta) =1$.  This relation can be expressed in the form of a
{\it unitarity triangle} in the complex $(\rho,\eta)$ plane, illustrated in
Fig.\ \ref{fig:ut}.  One learns its shape from various sources:

% This is Figure 8
\begin{figure}
\begin{center}
\includegraphics[height=2.2in]{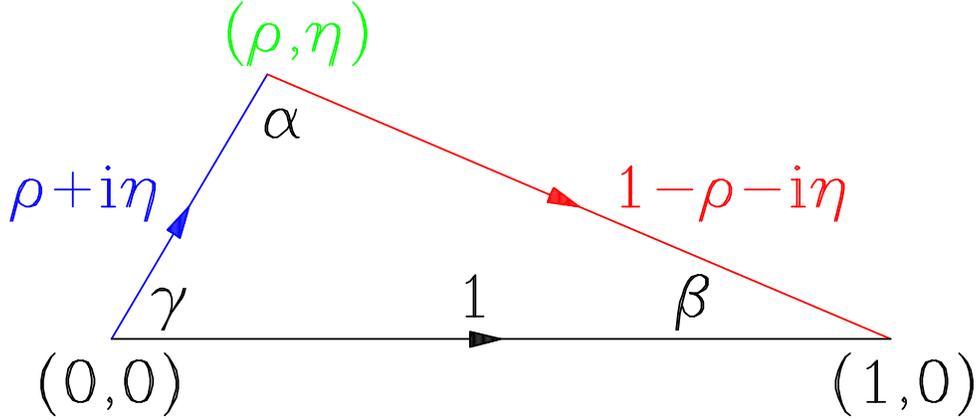}
\end{center}
\caption{The unitarity triangle.
\label{fig:ut}}
\end{figure}

\begin{itemize}

\item $B$--$\overline{B}$ mixing constrains $|V_{td}|$ and hence
$|1 - \rho - i \eta|$.

\item Charmless $B$ decays provide information on $|V_{ub}|$ and hence
$|\rho - i \eta| = (\rho^2+\eta^2)^{1/2}$.

\item $CP$-violating mixing in neutral kaon decays constrains Im$|V_{td}^2|$,
as mentioned, and hence provides information on $\eta (1-\rho)$.

\end{itemize}

What is remarkable is that all of these (and many other) constraints give a
consistent picture of the allowed $(\rho,\eta)$ region (see, e.g., Ref.\
\cite{Smith:2006}).  The non-zero phases, and the observation of large
$B$--$\overline{B}$ mixing in 1987 \cite{Albrecht:1987dr}, suggested that
$CP$ violation in $B$ meson decays would be {\it large}, as compared with
effects of order $10^{-3}$ in $K^0$ decays.

\section{NEUTRAL B'S: MIXING AND CP VIOLATION}

\begin{figure}
\begin{center}
\includegraphics[height=2.4in]{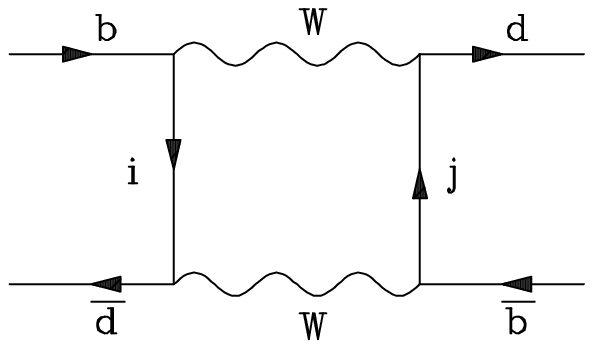}
\end{center}
\caption{Box diagram describing $\bo$--$\ob$ mixing.  Here $i,j = (u,c,t)$.
\label{fig:bmix}}
\end{figure}

The loop diagram of Fig.\ \ref{fig:bmix} allows $b \bar d \leftrightarrow d
\bar b$ transitions.  The matrix element of this operator at the quark level
is governed by $f^2_B B_B$, where $f_B$ is the $B$ meson decay constant and
the parameter $B_B$ describes the degree to which the vacuum intermediate state
dominates the $\Delta B = 2$ transition.  At present the best information on
$f_B^2 B_B$ comes from lattice QCD.  With $\Delta m (B^0) \simeq 0.5$
ps$^{-1}$, one is able to extract $|V_{td}|$ to only about 15\% from
$\bo$--$\ob$ mixing.

% This is Figure 9

One measures $|V_{ub}|$ through charmless semileptonic $B$ decays, constituting
only about 2\% of such decays.  One finds $|\rho - i \eta| \simeq 0.4 \pm 0.1$,
where I prefer to assign fairly conservative errors since the dominant
uncertainties in various methods of extracting $|V_{ub}|$ from data are
theoretical.

Strange $B$ ($B_s$--$\overline{B}_s$) mixing is governed by the same diagram as
in Fig.\ \ref{fig:bmix} but with $d \to s$.  Because $V_{ts} \simeq -V_{cb}$
is fairly well known, this measurement ends up providing information mainly
on $f_{B_s}^2 B_{B_s}$ and thus, through SU(3), on $f^2_B B_B$.  The mixing has
now been observed \cite{Abulencia:2006ze,D0mix}:  for example, the CDF
Collaboration finds $\Delta m_s = 17.77\pm0.10\pm 0.07$ ps$^{-1}$, leading to
the constraint $|V_{td}/V_{ts}| = 0.2060 \pm 0.0007~({\rm exp})^{+0.0081}
_{-0.0080}~({\rm theor})$ when one uses a value of $\xi \equiv (f_{B_s}
\sqrt{B_{B_s}}/f_B \sqrt{B_B}) = 1.21^{+0.047}_{-0.035}$ from lattice QCD
\cite{Okamoto}.

Some fairly precise information is now available about the angles of the
unitarity triangle.  The $CP$ asymmetry in $B^0 \to J/\psi K_S$ (comparing the
rate with that of $\overline{B}^0 \to J/\psi K_S$) measures $\sin 2 \beta =
0.674 \pm 0.026$, giving $\beta \simeq (21 \pm 1)^\circ$.  The $B_s$--%
$\overline{B}_s$ mixing just mentioned constrains $\gamma$ to be within a few
degrees of $\simeq 60^\circ$.  The asymmetric $e^+ e^-$ colliders and other
experiments seek to produce enough $B$'s to tell whether this picture is
self-consistent.  So far, it seems to be, but there are a few things to watch.
First, we sketch the way in which $B^0 \to J/\psi K_S$ provides information on
the angle $\beta$.

\subsection{CP asymmetry in $B^0 \to J/\psi K_S$}

There is a resonance $\Upsilon(4S)$ just above $B \bar B$ threshold.  If one
produces this resonance in $e^+ e^-$ collisions, the $\bo \ob$ pair is produced
in a correlated state.  As a result of $\bo$--$\ob$ mixing, one must
determine the relative proper times at which each $B$ meson decays in order to
properly ``tag'' the flavor of the produced $B$ meson.  The asymmetric electron
and positron  energies at the KEK and SLAC ``$B$ factories'' give the center of
mass a ``boost,'' allowing for easier detection of decay vertices.

\begin{figure}
\begin{center}
\includegraphics[height=4.7in]{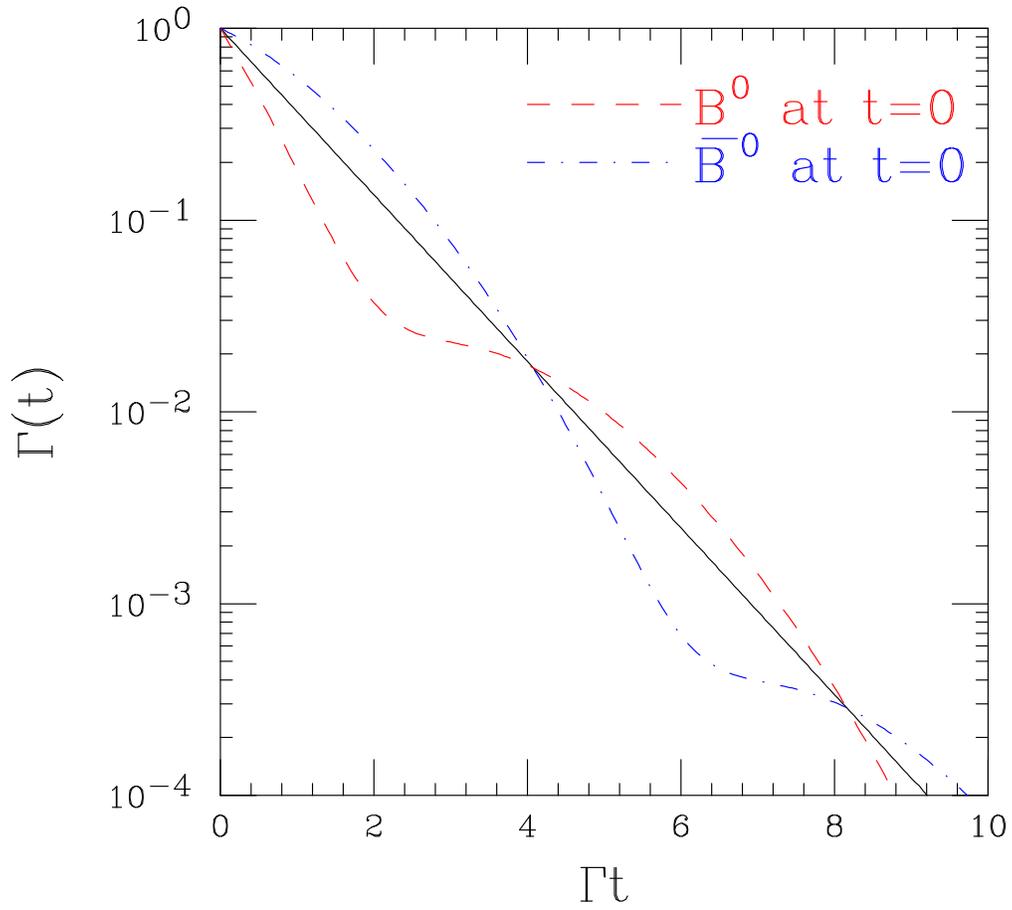}
\end{center}
\caption{Oscillations in $B^0(t)$ or $\ob(t)$ decays to $J/\psi K_S$.
\label{fig:oscs}}
\end{figure}

The decay rate of an initially-produced $\bo$ or $\ob$ is then given by
\beq
\Gamma(t) \left \{ \begin{array}{c}B^0_{t=0} \\
\ob_{t=0} \end{array} \right \} = e^{- \Gamma t}
[1 \mp \sin(2 \beta) \sin \Delta m t ]
\eeq
as a function of time, as illustrated in Fig.\ \ref{fig:oscs}.  Here the decay
rate is $\Gamma \simeq 0.65 \times 10^{12}$ s$^{-1}$, while the mixing
amplitude is $\Delta m \simeq 0.5 \times 10^{12}$ s$^{-1}$.  The first term
describes direct decay to $J/\psi K_S$, while the second describes decay
through mixing.  The $\bo$--$\ob$ mixing amplitude has a phase $2 \beta$.
The time-integrated decay asymmetry
\beq
\frac{\Gamma(\ob_{t=0} \to J/\psi K_S) - \Gamma(B^0_{t=0} \to J/\psi K_S)}
{\Gamma(\ob_{t=0} \to J/\psi K_S) + \Gamma(B^0_{t=0} \to J/\psi K_S)}
\eeq
would be maximal (= $\frac{1}{2} \sin (2 \beta) \simeq 0.34$)
if $\Delta m = \Gamma$; it is 97\% of that.

% This is Figure 10

\subsection{Time-dependent CP asymmetries}

% This is Figure 11
\begin{figure}
\begin{center}
\includegraphics[height=1.8in]{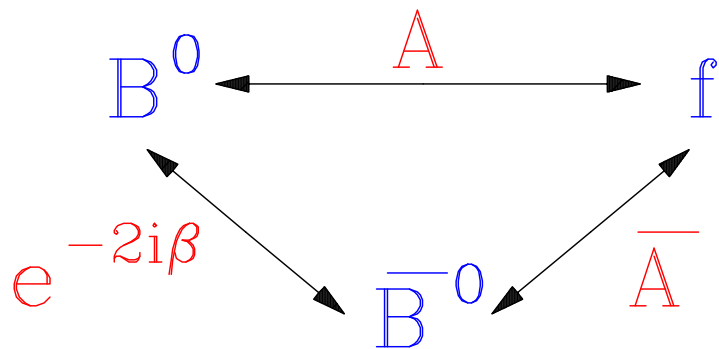}
\end{center}
\caption{Interference of decay and mixing in $B^0$ decays to a final state
$f$.
\label{fig:decmix}}
\end{figure}

The decay of a $\bo$ to a CP eigenstate $f$ can proceed either directly, via
the amplitude $A$ shown in Fig.\ \ref{fig:decmix}, or via mixing with a $\ob$,
followed by the decay $\ob \to f$ described by the amplitude $\bar A$.  The
resulting time-dependence in the most general case has the form
\beq \label{eqn:tdep}
\Gamma[B^0(t) \to f] \sim e^-\Gamma t~[ \cosh(\Delta \Gamma t/2)
- D~\sinh(\Delta \Gamma t/2) + C \cos(\Delta m t) - S \sin(\Delta m t)]
\eeq
with $C^2 + D^2 + S^2 = 1$, where 
\beq
\lambda \equiv e^{-2 i \beta}\frac{\bar A}{A}~,~~
S \equiv \frac{2 {\rm Im} \lambda}{1 + |\lambda|^2}~,~~
C \equiv \frac{1 - |\lambda|^2}{1 + |\lambda|^2}~,~~
D \equiv \frac{2 {\rm Re} \lambda}{1 + |\lambda|^2}~.
\eeq
These relations are reminiscent of Eqs.\ (1) and (2) for nonleptonic hyperon
decay involving the two complex amplitudes ($s,p$), or Stokes parameters
$(I,U,V,Q)$  describing radiation
polarization amplitudes ($E_\parallel,E_\perp$):
\beq
D \Leftrightarrow \alpha \Leftrightarrow U/I~,~~ 
S \Leftrightarrow \beta \Leftrightarrow V/I~,~~
C \Leftrightarrow \gamma \Leftrightarrow Q/I~.
\eeq
For the $B^0$, $\Delta \Gamma$ is small; so one measures $S,~C$
with $S^2 + C^2 \le 1$.  When $A$ is dominated by a single weak amplitude one
has
\beq
|\lambda| = 1~~,~~~C=0~~,~~~
S = \pm \sin[2 \beta + {\rm Arg}(A/\bar A)]~~.
\eeq

\subsection{Strange penguins?}

\begin{figure}
\begin{center}
\includegraphics[height=1.8in]{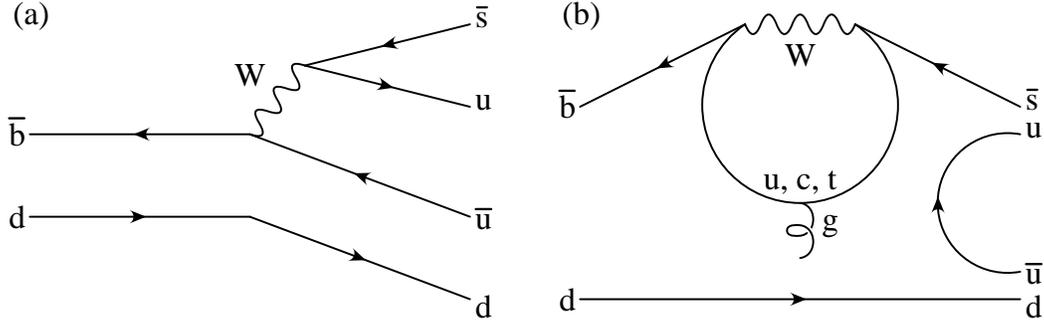}
\end{center}
\caption{Diagrams describing $B^0 \to \bar s u \bar u d$ decays.  (a) Tree;
(b) ``penguin.''
\label{fig:treepen}}
\end{figure}

A number of $B$ decay processes involving the virtual transition $\bar b \to
\bar s$ appear to be dominated by the ``strange penguin'' amplitude.  Diagrams
illustrating contributions to such decays are shown in Fig.\ \ref{fig:treepen}.
The name arose because the loser of a darts game in a pub near CERN in 1977
had agreed to use ``penguin'' in his next paper \cite{Ellis:1977uk}.

% This is Figure 12

For $B$ decays dominated by the $\bar b \to \bar s$ penguin, one expects the
coefficient $S$ of the $\sin \Delta m t$ decay rate modulation in Eq.\
(\ref{eqn:tdep}) to be $\pm \sin 2 \beta = 0.674\pm0.026$ as for $B^0 \to
J/\psi K_S$, where the $\pm$ sign denotes minus the CP eigenvalue of the final
state.  The observed values of $\sin(2 \beta)_{\rm eff}$ for many such
processes fall a bit below this value, as illustrated in Table \ref{tab:bspeng}
\cite{Hazumi}.  If deviations are due to new physics, should they be the same
in each case?

% This is Table II
\begin{table}
\caption{Values of $\pm S = \sin(2 \beta)_{\rm eff}$ for some $B$ decays
dominated by the $\bar b \to \bar s$ penguin amplitude.
\label{tab:bspeng}}
\begin{center}
\begin{tabular}{c c c c} \hline \hline
Final state & BaBar (SLAC) & Belle (KEK) & Average \\ \hline
$K_S \pi^0$ & $0.33\pm0.26\pm0.04$ & $0.33\pm0.35\pm0.08$ &
 $0.33 \pm 0.21$ \\
$K_S \eta'$ & $0.58\pm0.10\pm0.03$ & $0.64\pm0.10\pm0.04$ &
 $0.59 \pm 0.08$ \\
$K_S \phi$  & $0.12\pm0.31\pm0.10$ & $0.50\pm0.21\pm0.06$ &
 $0.39 \pm 0.18$ \\
\hline \hline
\end{tabular}
\end{center}
\end{table}

\section{UNFINISHED BUSINESS}

In the 110 years since Becquerel, experiment and theory have made great
strides in weak-interaction physics.  Rather than concluding, let me indicate
some questions for the future.

\subsection{$B$ decays involving $\bar b \to \bar s$ ``penguin'' diagrams}

One expects CP asymmetry parameters $\pm S = \sin(2 \beta)_{\rm eff} =
0.674 \pm 0.026$ in $B^0 \to K^0 \pi^0$, $B^0 \to \eta' K^0$, $B^0 \to \phi
K^0$.  These and other related $\bar b \to \bar s$ penguin
processes give an average $\sim 0.52 \pm 0.05$
\cite{Hazumi}, $2.6 \sigma$ below the expected value.  Standard Model
deviations from the nominal value have been calculated or bounded and are
expected to be small, less than 0.05 in many explicit calculations and less
than about 0.1 under very general circumstances \cite{Gronau:2006qh}.  We are
watching the situation with interest.

\subsection{Electroweak theory requires a Higgs boson}

Current thinking puts it just above the reach of recently
terminated LEP experiments.  The CERN Large Hadron Collider and possibly the
Fermilab Tevatron will have a shot at it.

\subsection{Pattern of quark masses and mixings}

Quark masses and mixings probably originate from the same physics, but there
seem to be no good ideas for understanding their pattern.  This is a central
question facing particle physics, and the community's inability to solve it has
led to a widespread feeling that it might be a question with no fundamental
answer, such as the radii of the planetary orbits.  I do not share this
pessimism.  The emerging pattern of neutrino masses and mixings will probably
provide important clues.

\section*{ACKNOWLEDGMENTS}
I would like to thank Jim Cronin for uncovering the nature of the
weak interaction through his studies of strange particle decays; for helping to
prove the importance of quarks via his experiments on particle production at
high transverse momenta; for his exciting explorations of cosmic rays at the
highest energies; and for making the University of Chicago such a wonderful
place to do physics.  I am grateful to Jim Pilcher for extending the invitation
to speak at this symposium and for helpful comments.
This work was supported in part by the United States Department of Energy
through Grant No.\ DE FG02 90ER40560.

\end{document}